\documentclass[11pt]{article}
\usepackage{a4wide}                       
\usepackage[dvips]{graphicx}
\usepackage{epsf}                        
\usepackage{latexsym}                    
\usepackage{amsfonts}                    
\usepackage{amssymb}                     

\def\be{\begin{equation}}
\def\ee{\end{equation}}
\def\bea{\begin{eqnarray}}
\def\eea{\end{eqnarray}}

\def\part{\partial}

\def\R{\ensuremath{\mathbb{R}}}

\def\makeatletter{\catcode`\@=11}
\makeatletter
\def\mathbox#1{\hbox{$\m@th#1$}}%
\def\math@ccstyles#1#2#3#4#5#6#7{{\leavevmode
      \setbox0\mathbox{#6#7}%
      \setbox2\mathbox{#4#5}%
      \dimen@ #3%
      \baselineskip\z@\lineskiplimit#1\lineskip\z@
      \vbox{\ialign{##\crcr
             \hfil \kern #2\box2 \hfil\crcr
             \noalign{\kern\dimen@}%
             \hfil\box0\hfil\crcr}}}}
\def\mathaccstyles{\math@ccstyles\maxdimen}
\def\maththroughstyles{\math@ccstyles{-\maxdimen}}
\def\unity%
 {\maththroughstyles{.45\ht0}\z@\displaystyle {\mathchar"006C}\displaystyle 1}
\begin{document}

\rightline{FFUOV-08/10}
\rightline{CERN-TH-PH-2008-185}
\vspace{1.5truecm}

\centerline{\LARGE \bf Confinement and Non-perturbative Tachyons}
\vspace{.5cm}
\centerline{\LARGE \bf in Brane-Antibrane Systems}
\vspace{1.3truecm}

\centerline{
    {\large \bf Norberto Guti\'errez${}^{a,}$}\footnote{E-mail address:
                                   {\tt norberto@string1.ciencias.uniovi.es}} 
    {\bf and}
   {\large \bf Yolanda Lozano${}^{a,b,}$}\footnote{E-mail address:
                                  {\tt ylozano@uniovi.es}}
    }
                                                            
\vspace{.4cm}

\centerline{{\it ${}^a$Departamento de F{\'\i}sica,  Universidad de Oviedo,}}
\centerline{{\it Avda.~Calvo Sotelo 18, 33007 Oviedo, Spain}}

\vspace{.4cm}
\centerline{{\it ${}^b$Theory Group, Physics Division, CERN,}}
\centerline{{\it CH-1211 Geneva 23, Switzerland}}

\vspace{1truecm}

\centerline{\bf ABSTRACT}
\vspace{.5truecm}

\noindent
We present a worldvolume effective action suitable for the study of the
confined phase of a $(Dp,\bar{Dp})$ system at weak coupling. We identify the mechanism by which the fundamental string  arises from this action as a confined electric flux string when the $Dp$ and the 
$\bar{Dp}$ annihilate. We construct an explicit dual action, more suitable for the
study of the strong coupling regime, and show that it realizes a generalized
Higgs-St\"uckelberg phase for the (relative) $(p-2)$-form field dual to the 
(overall) BI vector. This is the mechanism put forward by Yi and collaborators based on duality arguments in order to explain the breaking of the overall $U(1)$ gauge group at strong coupling. Indeed,
 in our dual description 
the Goldstone boson is a $(p-3)$-form magnetically charged with
respect to the 
overall BI vector field. This indicates that 
the condensing tachyonic objects originate from open $D(p-2)$-branes stretched between the
brane and the anti-brane. 
Our results  provide an explicit realization of the breaking of the overall $U(1)$ gauge
group perturbatively,
 in a way consistent
 with the duality symmetries of String Theory.\\
\\

\newpage

\section{Introduction}

$D\bar{D}$ systems have been widely used in the literature in the study of string theory in time dependent backgrounds (see \cite{Sen4} for a review), and more recently in the study of chiral symmetry breaking in holographic models of QCD \cite{CKP,BSS,DN,DN2,SS}.
It is well-known that the superposition of a D$p$-brane and an anti-D$p$-brane 
constitutes a non-BPS
system whose instability manifests itself in the existence of a complex
tachyonic mode in the open
strings stretched between the pair \cite{Sen4}. If when the tachyon rolls down to its true minimum its phase acquires a winding number,  because of its coupling to the relative U(1) vector field a magnetic vortex soliton is created. This vortex solution carries D$(p-2)$-brane charge,
 as inferred
from the coupling in the Chern-Simons action of the
($Dp$,$\bar{Dp}$):
\begin{equation}
\int_{\R^{p,1}}C_{p-1}\wedge (dA-dA^\prime)\, ,
\end{equation}
\noindent where $C_{p-1}$ stands for the RR $(p-1)$-form potential and
$A$  and $A^\prime$  for the Born-Infeld vector
fields on the brane and anti-brane. 
Charge conservation therefore implies that a $D(p-2)$-brane
is left as a topological soliton\footnote{This observation  can be made more explicit by showing that the worldvolume theory on the vortex solution is given by the DBI action on a $D(p-2)$-brane \cite{Sen3,HN}. See also \cite{HKLM}.}.
In this process the relative $U(1)$
vector field acquires a mass through the Higgs mechanism by eating the phase of the
tachyonic field, and is removed from the low energy spectrum. The 
overall U(1) vector field, under which the tachyon is neutral, 
remains however unbroken, posing a puzzle \cite{Sred,Witten,Yi}.

It was suggested in \cite{Yi}, based on the duality relation between the Type IIA superstring 
and M-theory, that the overall $U(1)$ is in the confined phase. The suggested
mechanism for this confinement is a dual Higgs mechanism in which magnetically charged 
tachyonic states 
associated to open $D(p-2)$-branes stretched
between the $Dp$ and the $\bar{Dp}$ condense. 
Evidence for such a situation comes from the
M-theory description of a $(D4,\bar{D4})$ system.

The superposition of a $D4$ and a $\bar{D4}$
is described in M-theory as an $(M5,\bar{M5})$ pair wrapped in the eleventh
direction. The open strings that connect the $D4$ and the $\bar{D4}$
are realized as open M2-branes wrapped in the eleventh direction and stretched between the
$M5$ and the $\bar{M5}$. These M2-branes must contain as well a complex
tachyonic excitation. Since the tachyon condensing charged object is in this case extended (a tachyonic worldvolume string) there are no ways to describe quantitatively this type of mechanism.
However, duality with the Type IIA superstring implies that whatever this mechanism is the
condensation of
this tachyonic mode should
 be accompanied by a non-trivial magnetic flux, in this case of the relative
antisymmetric tensor field
in the worldvolume of the $(M5,\bar{M5})$. This magnetic flux generates charge with
respect to the 3-form potential of eleven dimensional supergravity,
as inferred from the coupling in the 
$(M5,\bar{M5})$ Chern-Simons action\footnote{Here ${\hat C}_3$ stands for the 3-form of eleven dimensional supergravity and
${\hat A}_{2}$ and ${\hat A}_2^{\prime}$ for
the worldvolume 2-form fields on the $M5$  and the
$\bar{M5}$. Note that ${\hat
  A}_{2}$ (self-dual) and ${\hat A}_{2}^{\prime}$ (anti-self-dual)
combine to give an unrestricted relative 2-form field \cite{Yi}.}
\begin{equation}
\label{Chat}
\int_{\R^{1,5}}{\hat C}_{3}\wedge (d{\hat A}_{2}-d{\hat A}_{2}^{\prime})
\, .
\end{equation}

\noindent An M2-brane would then emerge as the remaining topological soliton.

Let us suppose that one performs now the reduction from M-theory 
along a worldvolume direction of the $(M5,\bar{M5})$ 
transverse to the stretched M2-branes \cite{Yi}. In this case a $(D4,\bar{D4})$ system is obtained in which
tachyonic D2-branes are stretched between the $D4$ and the $\bar{D4}$. 
Again, if this
tachyonic mode condenses in a vortex-like configuration, $B_{2}$-charge will be induced in the system, 
as the reduction from the previous coupling along a worldvolume direction transverse to the stretched M2-branes shows
\begin{equation}
\label{F1}
\int_{\R^{1,4}}B_{2}\wedge (dA_{2}-dA_{2}^{\prime})\, ,
\end{equation}
\noindent where now
$A_{2}$ and $A_{2}^{\prime}$ are associated to open
D2-branes ending on the $D4$ and the $\bar{D4}$. A fundamental string would then arise as the remaining topological soliton.

Note that in this case the
Higgs mechanism is intrinsically non-perturbative, given that this description
emerges after interchanging two
compact directions in M-theory. Indeed, the coupling (\ref{F1}) shows
that the worldvolume 
dynamics of the $(D4,\bar{D4})$ system is governed by the 2-form gauge
fields dual in the five dimensional worldvolume
to the BI vector fields. These fields couple in the worldvolume with inverse coupling, and are therefore more adequate to describe the strong coupling regime of the system.

Therefore, qualitatively the duality between Type IIA and M-theory predicts the occurrence
of both the perturbative and non-perturbative Higgs mechanisms for the
$(D4,\bar{D4})$ system. The same conclusion can be reached for arbitrary
$(Dp,\bar{Dp})$ systems by T-duality arguments \cite{Yi}.
Applying T-duality to the coupling (\ref{F1}) along $(p-4)$ transverse 
directions\footnote{Or along a spatial direction of the stretched D2-brane if $p< 4$.} one gets
\begin{equation}
\label{F12}
\int_{\R^{1,p}}B_{2}\wedge (dA_{p-2}-dA_{p-2}^{\prime})\, .
\end{equation}
This coupling indicates that the fundamental string would arise as a topological soliton in a 
dual Higgs mechanism \cite{Rey} in which magnetically charged tachyonic states associated to open
$D(p-2)$-branes stretched between the $Dp$ and the $\bar{Dp}$
condensed\footnote{When $p=3$ this is exactly the S-dual picture of the
  creation of a $D1$-brane as a vortex in a $(D3,\bar{D3})$ system \cite{Yi}.}.
In terms of the original variables this would translate into confinement of the overall 
$U(1)$, given that due to the opposite orientation of the
$\bar{Dp}$-brane the relative $(p-2)$-form field 
is dual in the $(p+1)$-dimensional worldvolume to the overall
BI vector field. Therefore, its
localized magnetic flux at strong coupling translates into a confined overall $U(1)$
electric flux at weak coupling. 

The explicit action that describes the dual Higgs
mechanism at strong coupling has not been constructed in the
literature, 
although some
qualitative 
arguments pointing at particular couplings have been given \cite{Yi,BHY,GHY}.
In any case, as we have mentioned, this mechanism is intrinsically
non-perturbative, and this makes this description
highly heuristic.

A related crucial question which was first addressed in
\cite{Seneffac,Sen2,BHY,GHY} 
is the possibility of describing both the perturbative and the
non-perturbative Higgs mechanisms simultaneously at weak coupling. 
Starting with Sen's action \cite{Seneffac,Sen2} reference
\cite{GHY} studied the Hamiltonian classical dynamics of the $(Dp,\bar{Dp})$
system, and showed that
it describes a massive relativistic string
fluid. The possibility of describing the region of vanishing tachyonic potential 
in terms of the $(p-2)$-form fields dual to the BI
vector fields was also addressed\footnote{This idea was also put forward in \cite{BHY} in
  the 2+1 dimensional case.} and although the explicit dual action was
not constructed it was argued that the dual Higgs mechanism proposed in \cite{Yi} could 
be realized explicitly if this action was the one associated to an Abelian Higgs model for
the relative $(p-2)$-form dual field.  The
fundamental string would then arise as a Nielsen-Olesen solution.  In this construction, however, the
$(p-3)$-form field playing the role of the Goldstone boson associated with
the dual magnetic objects did not have a clear string theory
origin. 

One of the results that we will present in 
this paper will be the construction of the explicit dual action 
describing the strongly coupled dynamics of the $(Dp,\bar{Dp})$ system in terms of the
$(p-2)$-form dual potentials and a $(p-3)$-form Goldstone boson. The generalization of Sen's action to 
include tachyonic couplings in a $(Dp,\bar{Dp})$ system \cite{Pesando,KW,MZ,KL,TTU,AIO,LS,Szabo,JT2,Sen3,Garousi4,Garousi3,Garousi2,Garousi5} describes, to second order in $\alpha^\prime$,
an Abelian Higgs model in which the Abelian field is the relative BI
vector of the brane and the antibrane and the phase of the tachyon plays the role of the associated Goldstone
boson. We will show however that the dual of this action does not describe an Abelian Higgs model for the relative $(p-2)$-form potential, contrary to the expectation in \cite{GHY}. The explicit dual Abelian Higgs model will instead arise from a different generalization of Sen's action from which we will be able to describe the confining phase (for the overall U(1)) of the $(Dp,\bar{Dp})$ system at weak coupling.

The dualization of the four-dimensional Abelian Higgs
model is known since long ago \cite{Sugamoto}, motivated
by the study of the confining phases of four dimensional Abelian gauge theories in the context of Mandelstam-'t Hooft duality \cite{MT}. The dual action constructed by Sugamoto describes the confining phase of four dimensional vector fields in terms of a massive 2-form field theory which is
 an extension of the model for massive relativistic hydrodynamics of Kalb and Ramond \cite{KR}. This field theory allows a   
  quantized vortex solution similarly to the creation of the Nielsen-Olesen string in the Abelian Higgs model. 
The extension of Sugamoto's construction to arbitrary $d$-dimensional $p$-form Abelian Higgs models was carried out more recently in \cite{QT}, with the aim at describing the confining phases of $p$-form field theories in a generalization of Mandelstam-'t Hooft duality.
In this general case the
dual action describing the confining phase is a massive $(p+1)$-form field theory. 

In this paper we will develop on the work of \cite{QT} and 
we will extend the construction in \cite{Sugamoto} to the $(p+1)$-dimensional Abelian Higgs model that describes the Higgs phase (for the relative U(1)) of a $(Dp,\bar{Dp})$ system. As we will see the massive Abelian field of the 
Abelian Higgs model can still be dualized in the standard way into a massless $(p-2)$-form field once the phase of the tachyon is dualized into a $(p-1)$-form. We will show that the dual action is of  the type of the massive $(p-1)$-form field theories discussed in \cite{QT}. Furthermore, we will show that a 
$D(p-2)$-brane can emerge as a confined electric flux brane associated to the overall $(p-2)$-form dual field. The precise mechanism involved in this process is the Julia-Toulouse mechanism \cite{JT,QT}, which as we will see is the exact contrary of the more familiar Higgs mechanism. 

The construction of the dual action is therefore useful in order to identify the mechanism by which a $D(p-2)$-brane can emerge 
at strong
coupling after the 
annihilation of a $Dp$ and a $\bar{Dp}$. However,  it sheds no light on the issue of the unbroken overall $U(1)$, nor on the creation of the fundamental string, since it involves only the overall $(p-2)$-form potential, and this field is dual to the relative BI vector field. Indeed, inspired by Mandelstam-'t Hooft duality one expects that the dual action describes the creation of the $D(p-2)$-brane in dual variables, 
since it should provide an explicit
realization of the duality between the Higgs phase (for the relative U(1)), described at weak coupling by Sen's action, and the confinement phase (for the overall $(p-2)$-form field) at strong coupling.
The Higgs phase for the relative $(p-2)$-form gauge potential at strong coupling should instead be dual 
to the confining phase for the overall $U(1)$ at weak coupling.  

In this paper we will present a worldvolume effective
action suitable to describe perturbatively the dynamics of the $(Dp,\bar{Dp})$ system in the confining phase for the overall U(1). Developing on the work of \cite{QT} we will start in the phase in which the tachyon vanishes, 
the Coulomb phase, and show that the confining phase arises after
the condensation of $(p-3)$-dimensional topological defects which are interpreted as the
end-points of $D(p-2)$-branes. 
We will see that the fundamental
string emerges at weak coupling as a confined electric flux string after a 
Julia-Toulouse mechanism in which a 2-form
gauge field associated to the fluctuations of the topological defects 
eats the overall $U(1)$ vector field.
We will also show, following \cite{QT} closely, that the confined
phase for the original overall $U(1)$ vector field can be studied 
in the strong coupling regime as a 
generalized Higgs-St\"uckelberg phase for its dual $(p-2)$-form field. The explicit dual action
is given by an Abelian Higgs model for the relative $(p-2)$-form potential. In this description the condensing tachyonic objects are identified as $(p-3)$-branes that originate from the end-points of open $D(p-2)$-branes stretched between the $Dp$ and the $\bar{Dp}$.
The fundamental string then emerges as a topological soliton after the condensation of this tachyonic
mode through a dual
Higgs mechanism \cite{Rey}. Therefore, through this construction we can make explicit  the mechanism suggested in \cite{Yi}
for realizing non-perturbatively the confinement of the overall $U(1)$. 

As we have seen the $(Dp,\bar{Dp})$ system admits two types of topological defects: particles and $(p-3)$-branes. 
The first originate as the end-points of open strings and are therefore perturbative in origin. The second originate as the end-points of non-perturbative open $D(p-2)$-branes and can therefore only be described in terms of $D(p-2)$-brane degrees of freedom in the strong coupling regime. 
We have seen however that using Julia and Toulouse's idea we can incorporate these degrees of freedom in the perturbative action, and study the confining phase for the overall U(1). If we combine the effective actions describing the Higgs phase for the relative U(1) and the confining phase for the overall U(1) we will be able to describe perturbatively the breaking of both gauge groups. We will see that from this action both the $D(p-2)$-brane and the fundamental string are realized as solitons in the common $(p+1)$-dimensional worldvolume. The $D(p-2)$-brane arises after a Higgs mechanism involving the relative U(1), and the F1 after a Julia-Toulouse mechanism involving the overall U(1).

The organization of this paper is as follows. In section 2 we construct the dual of the Abelian Higgs model that describes the $(Dp,\bar{Dp})$ system at weak string coupling. We see that 
contrary to expectation in \cite{BHY} it does not describe an Abelian Higgs model for the dual relative $(p-2)$-form potential. 
The worldvolume field content of the dual action consists on a $(p-1)$-form, dual to the phase of the tachyon, and two $(p-2)$-form fields dual to the BI vectors. We show that the $(p-1)$-form can become massive by eating the overall dual $(p-2)$-form potential through the Julia-Toulouse mechanism, and that a $D(p-2)$-brane arises as a confined electric flux brane in this process. Therefore the Higgs phase for the relative BI vector is mapped onto the confining phase for the overall $(p-2)$-form field, with a $D(p-2)$-brane arising either as a vortex solution after the Higgs mechanism at weak coupling or as a confined electric flux brane after the Julia-Toulouse mechanism at strong coupling. In section 3 we present 
our candidate action for describing
the confining phase of the overall BI vector field at weak coupling. We show that from this action the fundamental string arises as a confined electric flux string after a Julia-Toulouse mechanism. 
In section 4 we construct the dual of this action and show that it realizes a generalized Higgs-St\"uckelberg phase for the relative $(p-2)$-form field. Therefore, the confining phase for the overall BI vector is mapped onto the Higgs phase for the relative $(p-2)$-form field, with a fundamental string arising either as a confined electric flux string after the Julia-Toulouse mechanism at weak coupling or as a generalized vortex solution after the Higgs mechanism at strong coupling. Section 5 is our Discussion section. Here we present the action 
from which we can describe
simultaneously the Higgs phase for the relative U(1) and the confinement phase for the overall U(1) at weak string coupling.

\section{The $(Dp,\bar{Dp})$ system in dual variables}

The effective action describing a brane-antibrane pair has been extensively studied in the literature
using different approaches \cite{Pesando,KW,MZ,KL,TTU,AIO,LS,Szabo,JT2,Sen3,Garousi4,Garousi3,Garousi2,Garousi5}. Although the complete action has not been derived from first principles it is known to satisfy a set of consistency conditions  \cite{Sen3}. 
It is
invariant under gauge transformations of the tachyon phase and the relative BI vector:
$\chi\rightarrow \chi+\alpha(x)$, $A^{-}\rightarrow A^{-}+d \alpha$, it reduces to the sum of the BI effective actions for the $Dp$ and the $\bar{Dp}$ 
for zero tachyon, and it gives rise
to the action for a non-BPS $Dp$-brane \cite{Seneffac,Garousin,BRWEP,Kluson}
when modded out by $(-1)^{F_L}$ \cite{Sen4}. In the context of our discussion in this paper this action describes the Higgs phase for the relative BI vector field.

In this paper we will work to second order in
$\alpha^\prime$, and take the RR potentials
$C_{p-3}, C_{p-5},\dots$ to zero. 
We will also ignore the tachyonic couplings to the $C_{p-1}$ RR-potential derived in \cite{KW,KL,Garousi2}.
Thus, our action represents a truncated version of the $(Dp,\bar{Dp})$ action that can be derived from the results in
\cite{Pesando,KW,MZ,KL,TTU,AIO,LS,Szabo,JT2,Sen3,Garousi4,Garousi3,Garousi2,Garousi5} \footnote{Note that in comparing with the boundary string field theory results \cite{KMM} there is the usual discrepancy by $2\log{2}$ in the kinetic term of the tachyon
\cite{KL,TTU,JT2}.}.
We will see however that it contains the relevant couplings for describing the most important aspects of the dynamics of the $(Dp,\bar{Dp})$ system, both in the Higgs and in the confining phases\footnote{Once it is extended as we do in next section in order to incorporate the non-perturbative degrees of freedom associated to the $(p-3)$-brane topological defects.}.

Our starting point is the action: 
\begin{eqnarray}
\label{Higgsinitial}
&&S(\chi,A)=\int d^{p+1}x\, \Bigl\{ e^{-\phi}
\Bigl(\frac12 F^++B_2\Bigr)\wedge *
\Bigl(\frac12 F^+ +B_2\Bigr)+\nonumber\\
&&+\frac14 e^{-\phi} F^-\wedge * F^-
+ |T|^2 (d\chi -A^-)\wedge *
(d\chi -A^-)
+ d|T|\wedge * d|T| -V(|T|)\nonumber\\
&&+C_{p-1}\wedge F^-\Bigr\}\, .
\end{eqnarray}
Here we have set $2\pi\alpha^\prime=1$,
$A^+$ and $A^-$ are the overall and relative BI vector fields: $A^+=A+A^\prime$, $A^-=A-A^\prime$, and the
complex tachyon is parametrized as $T=|T|e^{i\chi}$. $V(|T|)$ is the tachyon potential \cite{Sen2}, whose precise form will be irrelevant for our analysis. Finally,
the background fields $B_2$ and $C_{p-1}$
are implicitly pulled-back into the $(p+1)$-dimensional worldvolume of the $(Dp,\bar{Dp})$. 

The coupling $\int C_{p-1}\wedge F^-$ is the one that we discussed in the introduction. It shows that when the tachyon condenses in a vortex-like configuration a $D(p-2)$-brane is generated as a topological soliton \cite{Sen4}, since the associated localized $F^-$ magnetic flux generates $C_{p-1}$ charge.
 In this process the relative $U(1)$ vector field eats the
scalar field $\chi$, gets a mass and is removed from the low energy spectrum.
The overall
$U(1)$ vector field, under which the tachyon is neutral, remains unbroken,  but it
is believed to be confined \cite{Yi,Seneffac,Sen2,BHY}.

In this section we construct the dual of the action (\ref{Higgsinitial}), and show that it describes the confining phase for the $(p-2)$-form potential dual to the relative BI vector field, thus providing an explicit realization of Mandelstam-'t Hooft duality for the Abelian Higgs model associated to the $(Dp,\bar{Dp})$ system. We also discuss the mechanism by which the $D(p-2)$-brane arises as a confined electric flux brane.

\subsection{The duality construction}

Let us focus on the worldvolume dependence of the action (\ref{Higgsinitial})
on $A^{+}$, $A^{-}$ and the phase of the tachyon.
 Note that since $A^{-}$ is massive it cannot  be dualized in the standard way.  We can however 
 use the standard procedure to dualize the phase of the tachyon and $A^+$. These fields are dualized, respectively, into a $(p-1)$-form, $W_{p-1}$, and a $(p-2)$-form, that we denote by $A_{p-2}^-$ given that due to the opposite orientation of the antibrane the relative and overall gauge potentials should be interchanged under duality. The intermediate dual action that is obtained after these two dualizations are carried out is such that $A^-$ becomes massless\footnote{Up to a total derivative term.} and can therefore be dualized in the standard way into a $(p-2)$-form, which we denote as $A_{p-2}^+$ \footnote{Alternatively, one can use a generalization of the intermediate action presented in \cite{Sugamoto}, from which it is possible to dualize a massive Abelian 1-form field.}. 
 
The final dual action reads:
\begin{eqnarray}
\label{dualinitialHiggs}
&&\hspace{-0.2cm}S(W_{p-1}, A_{p-2})=\int d^{p+1}x\Bigl\{ e^\phi \Bigl( \frac12 F^{+}_{p-1}+W_{p-1}+C_{p-1}\Bigr)\wedge * \Bigl( \frac12 F^{+}_{p-1}+W_{p-1}+C_{p-1}\Bigr)\nonumber\\
&&\hspace{-0.3cm}+ \frac14 e^\phi F^{-}_{p-1}\wedge * F^{-}_{p-1}+\frac{1}{4|T|^2}dW_{p-1}\wedge * dW_{p-1}
+d|T|\wedge * d|T|-V(|T|)-B_2\wedge F^{-}_{p-1}\Bigr\}
\end{eqnarray}
with the explicit duality rules being given by:
\begin{eqnarray}
\label{A+A-}
\frac12 F^++B_2&=&\frac12 e^\phi * F^{-}_{p-1}\\
\frac12 F^-&=&e^\phi *\Bigl( \frac12 F^{+}_{p-1}+W_{p-1}+C_{p-1}\Bigr)\\
d\chi-A^-&=&\frac{1}{2|T|^2}(-1)^{p-1}*dW_{p-1}\, .
\end{eqnarray}
Here we see that the relative and overall gauge potentials are
interchanged, 
as expected due to the opposite orientation of the antibrane. Note that for $p=3$ our notation is ambiguous. When analyzing this particular case we will  use $A^+$ and $A^-$ to denote the BI vector fields and ${\tilde A}^+$, ${\tilde A}^-$ to denote the dual vector fields associated to open D-strings ending on the branes.

The action (\ref{dualinitialHiggs}) is an extension of the actions proposed in \cite{QT}
for describing the confining phases of field theories of
compact antisymmetric tensors. After we discuss these actions in some
detail in the next section it will become clear that (\ref{dualinitialHiggs})
describes the confining phase for the overall $(p-2)$-form dual
potential. 
This phase arises after the condensation of zero-dimensional
topological 
defects which originate from the end-points of open strings stretched 
between the brane and the antibrane. The interpretation of the low energy mode  $W_{p-1}$
is as describing the fluctuations of these defects, and is such that away from
the defects $W_{p-1}=dA_{p-2}^+$.

Note that the gauge invariance $\chi\rightarrow \chi+\alpha(x)$, $A^{-}\rightarrow A^{-}+d \alpha$
of the original action has been mapped under the duality transformation into
$W_{p-1}\rightarrow W_{p-1}+d\Lambda_{p-2}$, $
  A_{p-2}^+\rightarrow 
A_{p-2}^{+}-2\Lambda_{p-2}$. This symmetry can be
gauge fixed by absorbing 
$F_{p-1}^{+}$ into $W_{p-1}$, which becomes then massive. 
The overall $A_{p-2}^{+}$ gauge potential is then removed
from the low 
energy spectrum, through a mechanism that is the exact contrary of the 
Higgs mechanism. This is the Julia-Toulouse mechanism mentioned in the introduction. Thus, the Julia-Toulouse
 mechanism is identified as the mechanism responsible for the removal of the relative U(1) at strong coupling.
 However it
clearly 
sheds no light on the removal of $A^{+}$.

When comparing the action (\ref{dualinitialHiggs}) to the actions describing the confining phases of antisymmetric field theories presented in \cite{QT} one sees that the modulus of the tachyon plays the role of the density of condensing topological defects. In a way one can think of $|T|$ as an indicator of how unstable the system is. Since the instability in the confining phase is originated by the presence of the topological defects it is reasonable to expect a relation between both quantities. In the confining models of Quevedo and Trugenberger a consistency requirement is that the antisymmetric field theory in the Coulomb phase is recovered for zero density of topological defects. This is indeed satisfied by our action (\ref{dualinitialHiggs}) for vanishing tachyon, since the $|T|\rightarrow 0$ limit forces the condition that $W_{p-1}$ must be exact and can therefore be absorbed through a redefinition of $A^+$. The action is then reduced to the
action describing the $(Dp,\bar{Dp})$ system in the Coulomb phase, i.e. to  (\ref{Higgsinitial}) for zero tachyon. 

Finally, following the analysis in \cite{Sugamoto} we can see
that a $D(p-2)$-brane arises as 
a confined electric flux brane after the Julia-Toulouse mechanism. In order to see this explicitly we need however to recall some basic facts on the construction of \cite{Sugamoto}, so we will postpone this discussion till the end of next section.

In the next section we present our candidate action for describing
the confining phase for the overall $U(1)$ at weak coupling.
We show that the fundamental string arises from this action as a confined electric flux string. By direct generalization of this analysis we also show that the $D(p-2)$-brane arises as a confined electric flux brane from the action (\ref{dualinitialHiggs}) derived in this section.

\section{Confinement at weak string coupling}

In this section we present our candidate action for describing the dynamics
of the $(Dp,\bar{Dp})$ system in the confining phase. We use the results in \cite{QT},
where an action describing the confined
phase of field theories of compact antisymmetric tensors of arbitrary rank was derived.
We start by summarizing the qualitative points that are relevant for our construction,
to later concretize these ideas to the
$(Dp,\bar{Dp})$ system. 
The reader is referred to \cite{QT} for a more detailed discussion.

Quevedo and Trugenberger  made explicit in the framework of antisymmetric field theories
an old idea in solid-state physics due 
to Julia and
Toulouse \cite{JT}.
These authors argued that for a compact tensor field of rank $(h-1)$ in $(p+1)$-dimensions a 
confined phase might arise
after the condensation of $(p-h-1)$-dimensional topological defects\footnote{The
mechanism by which these defects originate is irrelevant for the nature of the confining phase.}.
The fluctuations of the continuous distribution
of topological defects generate a new low-energy
mode in the theory which can be described by a new $h$-form, 
$W_h$, such that away from the defects $W_{h}=dA_{h-1}$, where
$A_{h-1}$ is the original tensor field. The main idea is to extend the $h$-form
in the topological invariant term\footnote{$S_h$ is an $h$-dimensional sphere surrounding the defect on an $(h+1)$-dimensional
hyperplane perpendicular to it, and $\omega_h$ is an $h$-form which is
exact 
outside $S_h$.}
\begin{equation}
\int_{S_h} \omega_h
\end{equation}
to the whole $\R^{p+1}$ space-time. In this way the
$(p-h)$-form $J_{p-h}=*(d\omega_h)$, which is zero outside the defect, picks up delta-like
singularities at the locations of the topological defects and can describe the conserved fluctuations
of their continuous
distributions. Note that due to $J_{p-h}=*(d\omega_h)$ the new degrees of freedom are associated
only with the gauge invariant part of $\omega_h$. 

The effective action describing the confining phase of the
antisymmetric tensor field then depends 
on 
a gauge invariant combination of the antisymmetric tensor field, $A_{h-1}$,
and the extended $h$-form, $W_{h}$. This
combination is such that when the density of topological defects vanishes
the original action describing the antisymmetric tensor field theory in the Coulomb phase is recovered. 

As discussed in \cite{QT}, the finite condensate phase is a natural generalization of the confinement phase for a vector gauge field. For compact QED in four dimensions the induced static potential between a particle and an antiparticle is linear at large distances, identifying the monopole condensate phase as a confinement phase. This computation can be generalized to arbitrary $(h-1)$-forms in $d$ dimensions. In this case the leading term in the induced action is the $h$-dimensional hypervolume enclosed by the $(h-1)$-dimensional closed hypersurface to which the $(h-1)$-form couples. For a more detailed discussion on the confining properties of these actions see \cite{QT,EW}.

Given that the worldvolume theory of a $(Dp,\bar{Dp})$ system is a vector field theory, the 
results in \cite{QT} for $h=2$ can be applied to this case, with some obvious
modifications coming from the couplings to the background gauge potentials associated to the
closed strings. In this case the Coulomb phase is the phase with zero tachyon, and it is 
therefore described\footnote{To second order in $\alpha^\prime$ and for
$C_{p-3}=C_{p-5}=\dots=0$.}
by the Lagrangian:
\begin{equation}
\label{Coulombinitial}
L(A)= e^{-\phi}
\Bigl(\frac12 F^{+} +B_{2}\Bigr)\wedge *
\Bigl(\frac12 F^{+} + B_{2}\Bigr)+\frac14 e^{-\phi} F^{-}\wedge * F^{-}
+C_{p-1}\wedge F^{-}\, .
\end{equation}

Developping now on the ideas in \cite{QT} for the $(Dp,\bar{Dp})$ system we have that the
topological defects whose condensation will give rise to the confining phase are $(p-3)$-branes,
which originate in this case from the end-points of $D(p-2)$-branes stretched between
the $Dp$ and the $\bar{Dp}$. The new mode associated to the fluctuations of the defects is 
described by a 2-form, $W_{2}$, which will couple in the action through a gauge
invariant combination with the overall $U(1)$ vector field\footnote{One could in principle expect
that $W_{2}$ coupled to either combination of the $U(1)$ vector fields, but we will see 
that consistency with S- and T-dualities implies that it must couple only
to the overall vector field. This will allow ultimately to explain the puzzle of the unbroken
overall $U(1)$ through confinement.}.
The action should depend as well on the density of topological defects, such that
when this density vanishes the original action in the Coulomb phase, given by (\ref{Coulombinitial}),
is recovered. In the actions constructed in \cite{QT}
the density of topological defects entered as a parameter which was 
interpreted as a new scale in the theory. We will see however that in the $(Dp,\bar{Dp})$ case
duality implies that the density of topological defects must be a dynamical quantity, because
it is related to the modulus of the tachyonic excitation of the open $D(p-2)$-branes
in the dual Higgs phase. We will denote this field by 
$|\tilde{T}|$ and, moreover, we will use the duality with the Higgs phase to include in the action its 
kinetic and potential terms.

The action that we propose for describing the confining phase of the $(Dp,\bar{Dp})$ system is then given by:
\begin{eqnarray}
\label{confinitial}
S(W_{2},A)&=&\int d^{p+1} x \Bigl\{e^{-\phi}
\Bigl(\frac12 F^{+} +W_{2}+ B_{2}\Bigr)\wedge *
\Bigl(\frac12 F^{+} +W_{2}+ B_{2}\Bigr)
+\frac14 e^{-\phi} F^{-}\wedge * F^{-}+\nonumber\\
&&+\frac{1}{4|\tilde{T}|^2}
dW_{2}\wedge *dW_{2}+ 
d|\tilde{T}|\wedge * d|\tilde{T}| -V(|\tilde{T}|)+C_{p-1}\wedge F^{-}\Bigr\}\, .
\end{eqnarray}
This action has been constructed under four requirements.  
One requirement is gauge invariance, both under gauge transformations of the
BI vector fields and under 
$W_{2}\rightarrow W_{2}+d\Lambda_{1}$, which ensures that only the gauge invariant part
of $W_{2}$ describes a new physical degree of freedom. This transformation must be 
supplemented by $A^{+}\rightarrow A^{+}- 2\Lambda_{1}$, a symmetry that has to be gauge
fixed.
The second  is
relativistic invariance. The third requirement is that the original action 
describing the Coulomb phase must be recovered when $|\tilde{T}|\rightarrow 0$. Indeed, when 
$|\tilde{T}|\rightarrow 0$ we must have that
$dW_{2}=0$, so that $W_{2}=d\psi_{1}$ for some 1-form $\psi_{1}$. This form
can then be absorbed by $A^{+}$, and the original action (\ref{Coulombinitial}) is recovered.
These requirements were the ones imposed in \cite{QT}.  The $(Dp,\bar{Dp})$ system, being a string theory object, must also satisfy consistency with
the duality symmetries of string theory. The implications
of this requirement will become more clear when we show the duality between this action and the
action describing the Higgs phase for the dual $(p-2)$-form gauge field. It implies in particular
that $W_{2}$ must couple only to the overall $U(1)$ vector field.

Now, in (\ref{confinitial}) 
$F^{+}$ can be absorbed by $W_{2}$, fixing the gauge symmetry
\begin{eqnarray}
W_{2}&\rightarrow& W_{2}+d\Lambda_{1}\nonumber\\
A^{+}&\rightarrow& A^{+}-2\Lambda_{1}\, ,
\end{eqnarray}
and the action can then be entirely formulated 
in terms of $W_{2}$ and the relative vector field:
\begin{eqnarray}
\label{confinitial2}
S(W_{2},A^{-})&=&\int d^{p+1}x \Bigl\{e^{-\phi}
\Bigl(W_{2}+ B_{2}\Bigr)\wedge *
\Bigl(W_{2}+ B_{2}\Bigr)
+\frac14 e^{-\phi} F^{-}\wedge * F^{-}+\nonumber\\
&&+\frac{1}{4|\tilde{T}|^2}
dW_{2}\wedge *dW_{2}+ 
d|\tilde{T}|\wedge * d|\tilde{T}| -V(|\tilde{T}|)+C_{p-1}\wedge F^{-}\Bigr\}\, .
\end{eqnarray}
In this process the original gauge field $A^{+}$ has
been eaten by the new gauge field $W_{2}$, and has therefore been removed from
the low energy spectrum. This solves the puzzle of the unbroken overall $U(1)$ at weak string coupling through the Julia-Toulouse
mechanism, which,
as we have seen, is the exact opposite of the Higgs mechanism. Let us now see how the fundamental string arises from this action.

Consider first the $p=3$ case, which can be directly compared to the results in \cite{Sugamoto}.  In this case
the action (\ref{confinitial2}) is a generalization of the action proposed in \cite{Sugamoto} to describe the confining phase of a four dimensional Abelian gauge theory. We recall from the introduction that this action was constructed as the dual of the four dimensional Abelian Higgs model, and that it
allows a quantized electric vortex solution similar to the Nielsen-Olesen string. We see below that in our case this solution is identified as a fundamental string.

The construction of the vortex solution in \cite{Sugamoto} considers a non-vanishing 2-form vorticity source\footnote{In the construction in \cite{Sugamoto} the vorticity source is created by the phase component of the Higgs scalar of the original Abelian Higgs model. In our case it is created by the phase component of the tachyon field associated to open D-strings connecting the $D3$ and the $\bar{D3}$. This will become clear after the analysis in the next section.} along the $x^3$ axis:
\begin{equation}
V_e^{3}=n\delta(x^1)\delta(x^2)\, , \qquad V_e^i=0\,\, {\rm for} \,\, i=1,2\, , \qquad \vec{V}_b=0
\end{equation}
where the subindices $e$ and $b$ refer to the electric and magnetic components, and looks for a static and axially symmetric solution with the following assumptions:
\begin{equation}
\partial_0 e^3=\partial_0 |{\tilde T}|=0\, , \qquad e^3=e^3(r)\, , \qquad |{\tilde T}|=|{\tilde T}|(r)\, , 
\end{equation}
\begin{equation}
e^i=0 \,\, {\rm for}\,\, i=1,2\, , \qquad \vec{b}=0
\end{equation}
where $\vec{e}$ and $\vec{b}$ refer to the electric and magnetic components of $W_2$, and
$r=\sqrt{(x^1)^2+(x^2)^2}$. The solution that is found represents a static circulation of flow around the $x^3$ axis, 
and satisfies the quantization condition
\begin{equation}
\label{elecflux}
\int_{D_\infty}e^3 ds=2\pi n\, ,
\end{equation}
where $D_\infty$ is a large domain in the $(x^1,x^2)$ plane including the origin. This solution corresponds to the Nielsen-Olesen string in the original Higgs model. As expected, the magnetic flux quantization condition has been mapped under duality onto an electric flux quantization condition, given by (\ref{elecflux}). The reader is referred to \cite{Sugamoto} for a more detailed discussion. For arbitrary $p$ it is easy to find a similar, generalized, electric vortex solution with the same properties.

Let us now see that the confined electric flux string solution corresponds in the $(Dp,\bar{Dp})$ case to the fundamental string. In this case we have an additional coupling
\begin{equation}
\int B_2\wedge *W_2
\end{equation}
in the effective action (\ref{confinitial2}), which shows that the quantized electric flux generates $B_2$-charge in the system.
Charge conservation then implies that the remaining topological soliton is the fundamental string.

As mentioned in the previous section, the $D(p-2)$-brane arises from the strongly coupled confining action (\ref{dualinitialHiggs}) derived in that section in a very similar way. In this case the vorticity source is a $(p-1)$-form which is created by the phase of the tachyon field in the original action (\ref{Higgsinitial}). Note that in all the duality transformations that we have discussed in this paper we have ignored total derivative terms. Had we kept  these terms in the dualization of the action (\ref{Higgsinitial}) we would have found a coupling
$\int dW_{p-1}\wedge d\chi$ in the dual action. This coupling can be rewritten in terms of a vorticity source, $V_{p-1}=*dd\chi$, as
$\int W_{p-1}\wedge *V_{p-1}$, giving then the generalization to arbitrary dimensions of the vorticity coupling in \cite{Sugamoto}. Let us suppose that we fix now $\chi=n\theta$, where $\theta$ is the azimuthal angle 
in the $(x^{p-1},x^p)$ plane. For $n\neq 0$ $\theta$ is not well defined on the worldvolume of a $(p-2)$-brane, and therefore the vorticity source is  non-vanishing in this worldvolume. Taking then  $V_{p-1}^{012\dots p-2}=n\delta(x^{p-1},x^p)$ and zero otherwise, we can look for a static and axially symmetric solution with the assumptions
\begin{equation}
\partial_0 W_{p-1}^{012\dots p-2}=\partial_0 |T|=0\, ,\qquad W^{012\dots p-2}_{p-1}=W_{p-1}^{012\dots p-2}(r)\, ,\qquad
|T|=|T|(r)\, ,
\end{equation}
where $r=\sqrt{(x^{p-1})^2+(x^p)^2}$ and all other components of $W_{p-1}$ are taken to vanish. In this case the solution that is found represents a static circulation of flow around the $(p-2)$-brane, and satisfies the quantization condition
\begin{equation}
\label{elecflux2}
\int_{D_\infty}W_{p-1}^{01\dots p-2}ds=2\pi n\, .
\end{equation} 
The coupling
\begin{equation}
\int C_{p-1}\wedge *W_{p-1}
\end{equation}
in the dual effective action (\ref{dualinitialHiggs}) then implies that this confined electric flux brane corresponds to the $D(p-2)$-brane, since it shows that the quantized electric flux 
(\ref{elecflux2})   generates $C_{p-1}$-charge in the system.
Therefore, the $D(p-2)$-brane arises either as a magnetic vortex solution after the Higgs mechanism at weak coupling or as confined electric flux brane after the Julia-Toulouse mechanism at strong coupling.

In the next section we show that the action (\ref{confinitial}) can be made exactly equivalent to an
 action describing the Higgs phase for the dual relative $(p-2)$-form potential. We also show that, as expected, the fundamental string arises from this strongly coupled action as a generalization of the Nielsen-Olesen magnetic vortex solution.

\section{Confinement at strong string coupling: The dual Higgs mechanism}

Let us consider the action (\ref{confinitial}) describing the confining phase for the overall U(1) at weak string coupling. Inspired by Mandelstam-'t Hooft duality we expect that the dual of this action describes the Higgs phase for the $(p-2)$-form field dual to the overall BI vector. The dualization of the BI vector fields in (\ref{confinitial}) takes place in the standard way, given that they only couple through their derivatives. In turn, the 2-form $W_2$ is massive, but it can still be dualized in the standard way from the intermediate dual action that is obtained after dualizing the BI vector fields, in which it only couples through its derivatives. Let us call the dual of this form, a $(p-3)$-form, 
$\chi_{p-3}$. The final dual action reads:
\begin{eqnarray}
\label{Higgsdual}
&&S(\chi_{p-3},A_{p-2})=\int d^{p+1}x \Bigl\{e^\phi \Bigl(\frac12 F^{+}_{p-1}+C_{p-1}\Bigr)\wedge *
\Bigl(\frac12 F_{p-1}^++C_{p-1}\Bigr)
+\frac{1}{4} e^\phi F_{p-1}^{-}\wedge *
F_{p-1}^{-}\nonumber\\
&&+|{\tilde T}|^2 \Bigl( d\chi_{p-3}-A^{-}_{p-2}\Bigr)\wedge *
 \Bigl( d\chi_{p-3}-A^{-}_{p-2}\Bigr)+
d|{\tilde T}|\wedge * d|{\tilde T}| -V(|{\tilde T}|)-
B_{2}\wedge F_{p-1}^{-}\Bigr\}
\end{eqnarray}
and the explicit duality relations are given by
\begin{eqnarray}
\frac12 F^{-}&=&e^\phi *\Bigl(\frac12 F_{p-1}^{+}+
C_{p-1}\Bigr)\\
\frac12 F^{+}+W_2+B_2&=&\frac12 e^\phi *F_{p-1}^{-}\\
\frac12 dW_2&=&{|\tilde T|}^2(-1)^{p-1} *\Bigl(d\chi_{p-3}-A^{-}_{p-2}\Bigr)\, .
\end{eqnarray}
Notice that once again the overall and the relative gauge fields are interchanged. 

The action (\ref{Higgsdual}) describes an Abelian Higgs model for the relative $(p-2)$-form field, with the dual $(p-3)$-form $\chi_{p-3}$ playing the role of the associated Goldstone boson. The effective mass term reads
\begin{equation}
|{\tilde T}|^2 \Bigl(d\chi_{p-3}-A_{p-2}^{-}\Bigr)^2
 \end{equation}
 and it is gauge invariant  under $\chi_{p-3}\rightarrow \chi_{p-3}+\Lambda_{p-3}$, $A^-_{p-2}\rightarrow A^-_{p-2}+d\Lambda_{p-3}$. That a coupling of this sort could drive the dual Higgs mechanism was suggested in \cite{Yi,BHY,GHY} (see also \cite{Rey}) although it could not be explicitly derived
 from the action describing the Higgs phase at weak coupling, i.e. from Sen's action.
 In this paper we have seen that consistently with Mandelstam-'t Hooft duality the dual Abelian Higgs model arises from the action describing the confining phase at weak coupling. In the dual action (\ref{Higgsdual}) 
  the dual Goldstone boson $\chi_{p-3}$ is associated to the fluctuations of the 
 $(p-3)$-dimensional topological defects that originate from the end-points of the $D(p-2)$-branes stretched between the $Dp$ and the $\bar{Dp}$. This is consistent with the fact that this field 
 is the worldvolume dual of the field $W_{2}$, which was accounting for these fluctuations in
 the confining action (\ref{confinitial}).
 Moreover,  we can identify for $p=3$  
  the
 condensing Higgs scalar as the modulus of the tachyonic mode associated to open D-strings stretched between the $D3$ and the $\bar{D3}$. Indeed when $p=3$  the action (\ref{Higgsdual}) 
 reads\footnote{Here we have used tildes to denote the dual fields, as mentioned in section 2.}:
 \begin{eqnarray}
\label{HiggsdualD3}
&&L(\chi,A)=\int d^{p+1}x\Bigl\{ e^{\phi}
(\frac12 {\tilde F}^{+} + C_{2})\wedge *
(\frac12 {\tilde F}^{+} + C_{2})
+\frac14  e^{\phi} {\tilde F}^{-}\wedge * {\tilde F}^{-}\nonumber\\
&&+ |{\tilde T}|^2 (d{\tilde \chi} -{\tilde A}^{-})\wedge *
(d{\tilde \chi} -{\tilde A}^{-})+
d|{\tilde T}|\wedge * d|{\tilde T}| -V(|{\tilde T}|)
-B_{2}\wedge {\tilde F}^{-}\Bigr\}\, ,
\end{eqnarray}
i.e. it is the S-dual of the original action (\ref{Higgsinitial}) describing the perturbative Higgs phase of the $(D3,\bar{D3})$ system. This is an important consistency check for the actions that we have constructed, 
although strictly speaking S-duality invariance would only be expected for zero tachyon, i.e. when the system becomes BPS and the worldvolume field content is not expected to change at strong coupling.
Note that in this duality relation
the modulus of the perturbative tachyon is mapped into $|{\tilde T}|$, which can then be interpreted as the modulus of the tachyonic excitation associated to the open D-strings. Since ${\tilde \chi}$ has also an interpretation as the phase of the dual tachyon we can think of 
${\tilde T}=|{\tilde T}|e^{i{\tilde \chi}}$ as the complex tachyonic mode associated to the D-strings stretched between the $D3$ and the $\bar{D3}$.  For $p\neq 3$ $|{\tilde T}|$ plays formally the role of the modulus of a tachyonic excitation.
However, since the tachyonic condensing charged object is in this case a $(p-3)$-brane 
the phase of the tachyon is replaced by a $(p-3)$-form\footnote{Reference \cite{Yi} suggests a more concrete relation between the field $\chi_{1}$ for $p=2$ and the phase of the dual tachyon, by
imagining the relevant string field defined over a loop space as $e^{i\oint \chi_{1}}$. Imposing single-valuedness in the loop space would then imply $\oint_{\Sigma} d\chi_{1}=n$.}.
It would be interesting to clarify the precise way in which these fields arise as open $D(p-2)$-brane modes. 

Finally, let us discuss the way the fundamental string arises from the action (\ref{Higgsdual}) when the $Dp$ and the $\bar{Dp}$ annihilate. If the brane and the antibrane annihilate through a generalized Higgs-St\"uckelberg mechanism in which $A_{p-2}^{-}$ gets a mass by eating the Goldstone boson $\chi_{p-3}$, we have that, if the Goldstone boson acquires a non-trivial winding number:
\begin{equation}
\int_{\R^{p-1}} F_{p-1}^{-}=\oint_{S^{p-2}}A_{p-2}^{-}= \oint_{S^{p-2}}d\chi_{p-3}=2\pi n\, ,
\end{equation}
$B_{2}$-charge is induced in the configuration 
through the coupling in (\ref{Higgsdual})
\begin{equation}
\int_{\R^{p,1}}B_{2}\wedge F_{p-1}^{-}\, .
\end{equation}
Charge conservation therefore implies that after the annihilation a fundamental string is left as a topological soliton. Since in this process the relative $(p-2)$-form field is removed from the low energy spectrum, and this field is dual to the original overall $U(1)$, this solves the puzzle of the unbroken $U(1)$, through the mechanism suggested in \cite{Yi}  which is intrinsically non-perturbative.

\section{Discussion}

As we have seen, a $(Dp,\bar{Dp})$ system admits two types of topological defects: particles and $(p-3)$-branes. The first are perturbative in origin, while the second are non-perturbative. The combined electric and magnetic Higgs mechanisms introduce mass gaps to both U(1) vector potentials, being the only remnants $D(p-2)$-branes and fundamental strings, realized as solitons on the common $(p+1)$-dimensional worldvolume. The system is described perturbatively in terms of Sen's action, which incorporates the tachyonic degrees of freedom associated to the perturbative point-like defects. However, in order to incorporate the non-perturbative degrees of freedom associated to the $(p-3)$-dimensional topological defects one has to restrict to the strong coupling regime of the theory, where the degrees of freedom associated to these defects become perturbative. Even in this case, as we have seen, it is not obvious to account for the right fields describing the tachyonic excitations. We have seen in this paper that it is also possible to incorporate the non-perturbative degrees of freedom associated to the extended topological defects in the weak coupling regime, using Julia and Toulouse's idea. Essentially one introduces a new form which describes the fluctuations of these defects and imposes a set of consistency conditions based on gauge invariance and duality. In section 3 we have presented the weakly coupled action that  is formulated in terms of this new form and the $U(1)$ vector fields associated to the open strings. In fact, one can combine this action with Sen's action in order to incorporate the degrees of freedom associated to both the zero dimensional and extended topological defects, with the explicit combined action being given by:
\begin{eqnarray}
\label{completa}
S(\chi,W_{2},A)&=&\int d^{p+1} x \Bigl\{e^{-\phi}
\Bigl(\frac12 F^{+} +W_{2}+ B_{2}\Bigr)\wedge *
\Bigl(\frac12 F^{+} +W_{2}+ B_{2}\Bigr)
+\frac14 e^{-\phi} F^{-}\wedge * F^{-}+\nonumber\\
&&+|T|^2(d\chi-A^-)\wedge * (d\chi-A^-)+d|T|\wedge * d|T|
+\frac{1}{4|\tilde{T}|^2}
dW_{2}\wedge *dW_{2}+\nonumber\\ 
&&+d|\tilde{T}|\wedge * d|\tilde{T}| -V(|T|)-V(|\tilde{T}|)+
C_{p-1}\wedge F^{-}\Bigr\}\, .
\end{eqnarray}
This action describes both 
 the perturbative and the non-perturbative Higgs mechanisms simultaneously at weak coupling, and
it admits both a magnetic vortex solution, which by charge conservation is identified with the $D(p-2)$-brane, and an electric vortex solution, identified as the fundamental string. 

Finally, we would like to comment on two alternative mechanisms for recovering the fundamental string after $D\bar{D}$ annihilation that have been proposed in \cite{HKLM,KLS,KK} and in \cite{GHY,KY,Senlast,YY,GY}\footnote{See also \cite{GV,TT,DKKMMS,KMS}, in the framework of $c=1$ matrix models.}. In the first proposal \cite{HKLM,KLS,KK} the fundamental string emerges as a classical solution to Sen's action, with confinement being realized through the dielectric effect 
of \cite{KS}, with the tachyon potential playing the role of
the dielectric constant. The second proposal \cite{GHY,SenF1,GHY2,KY,Senlast,YY,GY} is based on the study of the description of the string fluid of \cite{GHY} in terms of closed strings. In this setup when there is a net electric flux the energy of the electric flux lines is associated to winding modes of fundamental strings.
These mechanisms are distinct to the one that we have proposed in this paper. In particular they do not seem to have a simple relation with the dual Higgs mechanism of 
\cite{Yi,BHY}.

\subsection*{Acknowledgements}

We would like to thank Fernando Quevedo for useful discussions.
N.G. would like to thank the IFT at Universidad Aut\'onoma de Madrid for hospitality while some parts of this work were done. Y.L. would like to thank the Theory Division at CERN for hospitality and support. The work of N.G. was supported by a FPU Fellowship from the Spanish Ministry of Education. This work has been partially supported by the CICYT grant MEC-06/FPA2006-09199 and by the European Commission FP6 program MRTN-CT-2004-005104, in which the authors are associated to Universidad Aut\'onoma de Madrid.


\begin{thebibliography}{99}

\bibitem{Sen4}
A. Sen, ``Tachyon dynamics in open string theory'', Int. J. Mod. Phys. A20 (2005) 5513, [arXiv:hep-th/0410103].

\bibitem{CKP}
R. Casero, E. Kiritsis, A. Paredes, ``Chiral symmetry breaking as open string tachyon condensation'', [arXiv:hep-th/0702155].

\bibitem{BSS}
O. Bergman, S. Seki, J. Sonnenschein, ``Quark mass and condensate in HQCD'', arXiv:0708.2839 [hep-th].

\bibitem{DN}
A. Dhar, P. Nag, ``Sakai-Sugimoto model, tachyon condensation and chiral symmetry breaking'', 
JHEP 0801 (2008) 055, arXiv:0708.3233 [hep-th].

\bibitem{DN2}
A. Dhar, P. Nag, ``Tachyon condensation and quark mass in modified Sakai-Sugimoto model'', arXiv:0804.4807 [hep-th].

\bibitem{SS}
T. Sakai, S. Sugimoto, ``Low energy hadron physics in holographic QCD'', Prog. Theor. Phys. 113 (2005) 843, [arXiv:hep-th/0412141]; ``More on a holographic dual of QCD'', Prog. Theor. Phys. 114 (2006) 1083 [arXiv:hep-th/0507073].

\bibitem{Sen3}
A. Sen, ``Dirac-Born-Infeld action on the tachyon kink and vortex'', Phys. Rev. D68 (2003) 066008, [arXiv:hep-th/0303057].

\bibitem{HN}
K. Hashimoto, S. Nagaoka, ``Realization of brane descent relations in effective theories'', Phys. Rev. D66 (2002) 026001, [arXiv:hep-th/0202079].

\bibitem{HKLM}
J.A. Harvey, P. Kraus, F. Larsen, E.J. Martinec,
``D-branes and strings as non-commutative solitons'',
JHEP 0007 (2000) 042, [arXiv:hep-th/0005031].


\bibitem{Sred}
M. Srednicki, ``IIB or not IIB'', JHEP 9808 (1998) 005,
[arXiv:hep-th/9807138].

\bibitem{Witten}
E. Witten, ``D-branes and K-theory'', JHEP 9812 (1998)
019, [arXiv:hep-th/9810188].

\bibitem{Yi}
P. Yi, ``Membranes from five-branes and fundamental strings from Dp-branes'', 
Nucl. Phys. B550 (1999) 214, [arXiv:hep-th/9901159].

\bibitem{Rey}
S. Rey, ``Higgs mechanism for Kalb-Ramond gauge field'',
Phys. Rev. D40 (1989) 3396.

\bibitem{BHY}
O. Bergman, K. Hori, P. Yi, ``Confinement on the brane'', 
Nucl. Phys. B580 (2000) 289,
[arXiv:hep-th/0002223].

\bibitem{GHY}
G. Gibbons, K. Hori, P. Yi, ``String fluid from unstable D-branes'', 
Nucl. Phys. B596 (2001) 136,
[arXiv:hep-th/0009061].

\bibitem{Seneffac}
A. Sen, ``Supersymmetric worldvolume action for non-BPS D-branes'',
JHEP 9910 (1999) 008, [arXiv:hep-th/9909062].

\bibitem{Sen2}
A. Sen, ``Universality of the tachyon potential'',
JHEP 9910 (1999) 008, [arXiv:hep-th/9911116].

\bibitem{Pesando}
I. Pesando, ``On the effective potential of the $Dp$-$\bar{Dp}$ system in Type II theories'',
Mod. Phys. Lett. A14 (1999) 1545, [arXiv:hep-th/9902181].

\bibitem{KW}
C. Kennedy and A. Wilkins, ``Ramond-Ramond couplings on brane-antibrane systems'',
Phys. Lett. B464 (1999) 206,
[arXiv:hep-th/9905195].

\bibitem{MZ}
J.A. Minahan, B. Zwiebach, ``Gauge fields and fermions in tachyon effective field theories'',
JHEP 0102 (2001) 034, [arXiv:hep-th/0011226].

\bibitem{KL}
P. Kraus, F. Larsen, ``Boundary string field theory of the $D\bar{D}$ system'', Phys. Rev. D63 (2001) 106004, [arXiv:hep-th/0012198].

\bibitem{TTU}
T. Takayanagi, S. Terashima, T. Uesugi, ``Brane-antibrane action from boundary string field theory'',
JHEP 0103 (2001) 019, [arXiv:hep-th/0012210].

\bibitem{AIO}
M. Alishahiha, H. Ita and Y. Oz, ``On superconnections and the tachyon effective action'',
Phys. Lett. B503 (2001) 181, [arXiv:hep-th/0012222]. 

\bibitem{LS}
N.D. Lambert, I. Sachs, ``On higher derivative terms in tachyon effective actions'', JHEP 0106
(2001) 060, [arXiv: hep-th/0104218].

\bibitem{Szabo}
R.J. Szabo, ``Superconnections, anomalies and non-BPS brane charges'', J. Geom. Phys. 43 (2002) 241, [arXiv:hep-th/0108043].

\bibitem{JT2}
N.T. Jones, S.-H.H. Tye, ``An improved brane anti-brane action from boundary superstring field theory and multi-vortex solutions'',
JHEP 0301 (2003) 012,
[arXiv:hep-th/0211180].

\bibitem{Garousi4}
M.R. Garousi, ``D-brane anti-D-brane effective action and brane interaction in open string channel'', JHEP 0501 (2005) 029, [arXiv:hep-th/0411222].

\bibitem{Garousi3}
M.R. Garousi, ``On the effective action of D-brane-anti-D-brane system'', arXiv:0710.5469 [hep-th].

\bibitem{Garousi2}
M.R. Garousi, E. Hatefi, ``On Wess-Zumino terms of brane-antibrane systems'', arXiv:0710.5875 [hep-th].

\bibitem{Garousi5}
M.R. Garousi, ``Higher derivative corrections to Wess-Zumino action of brane-antibrane systems'',
arXiv:0712.1954 [hep-th].





\bibitem{Sugamoto}
A. Sugamoto, ``Dual transformation in Abelian gauge theories'',
Phys. Rev. D19 (1979) 1820.

\bibitem{MT}
S. Mandelstam, ``Vortices and quark confinement in non-Abelian gauge theories'', Phys. Rep. C23 (1976) 237; 
G. 't Hooft, ``On the phase transition towards permanent quark confinement'', Nucl. Phys. B138 (1978) 1.

\bibitem{KR}
M. Kalb, P. Ramond, ``Classical direct interstring action'',
Phys. Rev. D9 (1974) 2273.


\bibitem{QT}
F. Quevedo, C. Trugenberger, ``Phases of antisymmetric tensor fields'',
Nucl. Phys. B501 (1997) 143,
[arXiv:hep-th/9604196].

\bibitem{JT}
B. Julia, G. Toulouse,  
J. Physique, Lett. 40 (1979) 396.

\bibitem{Garousin}
M.R. Garousi, ``Tachyon couplings on non-BPS D-branes and Dirac-Born-Infeld action'', Nucl. Phys. B584 (2000) 284, [arXiv:hep-th/0003122].

\bibitem{BRWEP}
E.A. Bergshoeff, M. de Roo, T.C. de Wit, E. Eyras, S. Panda, ``T-duality and actions for non-BPS D-branes'', JHEP 0005 (2000) 009, [arXiv:hep-th/0003221].

\bibitem{Kluson}
J. Kluson, ``Proposal for non-BPS D-brane action'', Phys. Rev. D62 (2000) 126003, [arXiv:hep-th/0004106].


\bibitem{KMM}
D. Kutasov, M. Mari\~no, G.W. Moore, ``Remarks on tachyon condensation in superstring field theory'', [arXiv:hep-th/0010108].

\bibitem{EW}
U. Ellwanger, N. Wschebor, ``Confinement with Kalb-Ramond fields'', 
JHEP 0110 (2001) 023, [arXiv:hep-th/0107196].


\bibitem{KLS}
M. Kleban, A. Lawrence, S. Shenker, ``Closed strings from nothing'',
Phys. Rev. D64 (2001) 066002, [arXiv:hep-th/0012081].

\bibitem{KK}
H. Kawai, T. Kuroki,  ``Strings as flux tube and deconfinement on branes in gauge theories'',
Phys.Lett. B518 (2001) 294, [arXiv:hep-th/0106103].


\bibitem{KS}
J. Kogut, L. Susskind, ``Vacuum polarization and the absence of free quarks in four dimensions'',
Phys. Rev. D9 (1974) 3501.

\bibitem{SenF1}
A. Sen, ``Fundamental strings in open string theory at the tachyonic vacuum'', J. Math. Phys. 42 (2001) 2844, [arXiv:hep-th/0010240].



\bibitem{GHY2}
G. Gibbons, K. Hashimoto, P. Yi, ``Tachyon condensates, Carrollian contraction of Lorentz group, and fundamental strings'', JHEP 0209 (2002) 061, [arXiv:hep-th/0209034].

\bibitem{KY}
O.K. Kwon, P. Yi, ``String fluid, tachyon matter and domain walls'', JHEP 0309 (2003) 003,
[arXiv:hep-th/0305229].

\bibitem{Senlast}
A. Sen, ``Open-closed duality at tree level'', Phys. Rev. Lett. 91 (2003) 181601,
[arXiv:hep-th/0306137].

\bibitem{YY}
H.U. Ye, P. Yi, ``Open/closed duality, unstable D-branes and coarse-grained closed strings'', 
Nucl. Phys. B686 (2004) 31,
[arXiv:hep-th/0402027].

\bibitem{GY}
M. Gutperle, P. Yi, ``Winding strings and decay of D-branes with flux'', JHEP 0501 (2005) 015,
[arXiv:hep-th/0409059].

\bibitem{GV}
J. McGreevy, H. Verlinde, ``Strings from tachyons: The c=1 matrix reloaded'', JHEP 0312 (2003) 054, [arXiv:hep-th/0304224].

\bibitem{TT}
T. Takayanagi, N. Toumbas, ``A matrix model dual of Type 0B string theory in two dimensions'', JHEP 0307 (2003) 064, [arXiv:hep-th/0307083].

\bibitem{DKKMMS}
M.R. Douglas, I.R. Klebanov, D. Kutasov, J. Maldacena, E. Martinec, N. Seiberg, ``A new hat for the c=1 matrix model'', [arXiv:hep-th/0307195].

\bibitem{KMS}
I.R. Klebanov, J. Maldacena, N. Seiberg, ``D-brane decay in two dimensional string theory'', JHEP 0307 (2003) 045, [arXiv:hep-th/0305159].















\end{thebibliography}
\end{document}